\documentstyle[emulateapj,psfig]{article}

\begin{document}
\submitted{Accepted for ApJ Letters}
\title{HST and VLA Observations of the H$_2$O Gigamaser Galaxy
TXS2226-184$^1$}

\footnotetext[1]{Based on observations with the NASA/ESA Hubble Space
Telescope, obtained at the Space Telescope Science Institute, which is
operated by AURA, Inc., under NASA contract NAS 5-26555 and on
observations made with the VLA operated by NRAO. The National Radio
Astronomy Observatory is a facility of the National Science
Foundation, operated under a cooperative agreement by Associated
Universities, Inc.}

\author{Heino Falcke\altaffilmark{2}, Andrew
S. Wilson\altaffilmark{3,4}, Christian Henkel\altaffilmark{2}, 
Andreas Brunthaler\altaffilmark{2},
James A. Braatz\altaffilmark{5}}
\altaffiltext{2}{Max-Planck-Institut f\"ur Radioastronomie, Auf dem H\"ugel 69,
D-53121 Bonn, Germany (hfalcke,chenkel,brunthal@mpifr-bonn.mpg.de)}
\altaffiltext{3}{Astronomy Department, University of Maryland, College Park,
MD 20742-2421 (wilson@astro.umd.edu)}
\altaffiltext{4}{Adjunct Astronomer, Space Telescope Science Institute}
\altaffiltext{5}{National Radio Astronomy Observatory, P.O. Box 2,
Green Bank, WV 24944 (jbraatz@nrao.edu)}

\begin{abstract}
We present HST/WFPC2 images in
H$\alpha$+[\ion{N}{2}]$\lambda\lambda$6548,6583 lines and continuum
radiation and a VLA map at 8 GHz of the H$_2$O gigamaser galaxy
TXS2226-184. This galaxy has the most luminous H$_2$O maser emission
known to date. Our red continuum images reveal a highly elongated
galaxy with a dust lane crossing the nucleus. The surface brightness
profile is best fitted by a bulge plus exponential disk model,
favoring classification as a highly inclined spiral galaxy
($i=70^\circ$). The color map confirms the dust lane aligned with the
galaxy major axis and crossing the putative nucleus. The
H$\alpha$+[\ion{N}{2}] map exhibits a gaseous, jet-like structure
perpendicular to the nuclear dust lane and the galaxy major axis. The
radio map shows compact, steep spectrum emission which is elongated in
the same direction as the H$\alpha$+[\ion{N}{2}] emission. By analogy
with Seyfert galaxies, we therefore suspect this alignment reflects an
interaction between the radio jet and the ISM. The axes of the nuclear
dust disk, the radio emission, and the optical line emission
apparently define the axis of the AGN. The observations suggest that
in this galaxy the nuclear accretion disk, obscuring torus, and large
scale molecular gas layer are roughly coplanar. Our classification of
the host galaxy strengthens the trend for megamasers to be found
preferentially in highly inclined spiral galaxies.
\end{abstract}


\section{Introduction}
In recent years a number of active galaxies have been found to have
powerful H$_2$O maser emission in their nuclei (e.g.~Braatz, Wilson,
\& Henkel 1994; 1996). It is known that the H$_2$O megamaser
phenomenon is associated with nuclear activity since all such
megamaser sources are in either Seyfert 2 or LINER nuclei. The
standard model for Seyfert galaxies involves a central engine (black
hole and accretion disk) producing ionizing radiation, and an
``obscuring torus'' which shadows the ionizing radiation into
bi-conical beams along its rotation axis (see Antonucci 1993 for a
review). This beaming is readily seen in some Seyferts as bi-conical
emission-line structures (e.g. Pogge 1989). Extended radio emission,
when present, is usually aligned with the emission-line gas
(e.g. Wilson \& Tsvetanov 1994). Detailed studies also indicate a
strong interaction between the radio ejecta and the optically visible
ionized gas (Capetti et al.~1996; Falcke et al.~1996; Falcke, Wilson,
\& Simpson 1998; Ferruit et al.~1999).

It appears reasonable to infer that the masers trace molecular
material associated with the obscuring torus or an accretion disk that
feeds the nucleus.  This notion was confirmed in great detail by VLBI
observations of the megamaser in NGC 4258 (Miyoshi et al. 1995;
Greenhill et al. 1995). The positions and velocities of the H$_2$O
maser lines show that the masing region is a thin disk in Keplerian
rotation around a central mass of $3.9\cdot10^7M_\odot$ at a distance
of $\approx$0.16 pc from that mass (Herrnstein et al. 1999).

Although plausible scenarios for the megamaser phenomenon exist
(e.g. Neufeld \& Maloney 1995), it is by no means clear how the
material which obscures the nucleus (the ``obscuring torus'') and
the masing disk are related. The masing disk may
be part of a geometrically thin, molecular accretion disk at smaller
radii than the torus, or   the thin, central plane of a thick
torus in which the column density is high enough for strong
amplification. Alternatively, the whole structure could be a warped thin
disk, so the masing gas might be misaligned with the central accretion
disk. The most straightforward picture consistent with current data
would, however, have the masing disk, obscuring torus and any more
extended molecular cloud distribution as one coherent accretion
structure feeding the central engine, with the ionized thermal and
non-thermal radio plasma roughly along the rotation axis.

We have therefore started a program to observe the narrow-line regions
(NLR) of all known megamaser galaxies with the Hubble-Space-Telescope
(HST) to establish this often suggested link between the molecular
disk responsible for the maser emission and the obscuring torus
responsible for the ionization cones. We are also obtaining continuum
color images to search for the obscuring material directly.

The most luminous known H$_2$O maser source is found in the galaxy
TXS2226-184\footnote{The name used by Koekemoer et al.~(1995) does not
follow the suggested and by now accepted convention used later in the Texas
survey (Douglas et al.~1996).} (IRAS F22265-1826; Koekemoer et
al. 1995), at a redshift of z=0.025 (luminosity distance D=101 Mpc
for $H_0=75$ km sec$^{-1}$ Mpc$^{-1}$ and $q_0=0.5$; in the images
0\farcs1 correspond to 46 pc). Koekemoer et al.~(1995) referred to
this object as a gigamaser in view of its   isotropic
luminosity in the 1.3 cm water line of 6100$\pm900 L_{\sun}$.  In this
paper, we present H$\alpha$+[\ion{N}{2}]$\lambda\lambda$6548,6583 and
broad-band continuum observations of TXS2226-184, obtained with the HST
and the VLA. Our results indeed show a linear H$\alpha$+[\ion{N}{2}]
structure along the radio axis and perpendicular to a dust lane. This
supports the connection between megamaser emission, dusty disk,
obscuring torus, and the narrow-line region discussed above. We also
classify the host galaxy as a spiral.

\section{Observations and Data Reduction}
\subsection{HST Observations}
TXS2226-184 was observed with the Planetary Camera (PC) on board the
HST (pixel scale is 0\farcs0455/pixel) in three filters: F814W (red
continuum); F547M (green continuum); and F673N (redshifted
H$\alpha$+[\ion{N}{2}]$\lambda\lambda$6548,6583). The total integration
times were 120 sec, 320 sec, and 1200 sec respectively, all exposures
being split into two or three integrations to allow cosmic ray
rejection. All observations were performed within one orbit on 1998
December 6.

\subsection{HST Data Reduction}
The images were processed through the standard Wide-Field and
Planetary Camera 2 (WFPC2) pipeline data reduction at the Space
Telescope Science Institute. Further data reduction was done in IRAF
and included: cosmic ray rejection, flux calibration, and rotation to
the cardinal orientation. The zero of magnitude for each continuum
filter was determined from the HST data handbook in the
VEGAMAG\footnote{A system in which Vega has magnitude zero in all HST
filters. The zero-points of the canonical Johnson-Cousins system
differ from the corresponding HST filters by up to 0.02 magnitudes for
closely matched filters and up to 0.2 mag for the rest.}
system. Sometimes we will refer to the red and green continuum filters
as I and V, respectively, even though F547M is not a good match to
Johnson-Cousins V; an error of 0.2 mag can be expected.  For the
continuum filters a constant background level was determined in an
emission-free region of the PC (to represent sky brightness) and
subtracted from the image. This correction is mainly important for
obtaining good color information in faint regions. The galaxy
continuum near the H$\alpha$+[\ion{N}{2}] line was determined by
combining the red and green continuum images, scaled to the filter
width of F673N and weighted by the relative offset of their mean
wavelengths from the redshifted H$\alpha$+[\ion{N}{2}] emission. The
continuum was then subtracted from the on-band image to obtain an
image of H$\alpha$+[\ion{N}{2}]. We did not apply any shifts between
the images because they were all taken within one orbit and at the
same position on the PC chip. From the two broad-band images, we
constructed a color map by dividing the green by the red filter image,
including only pixels where the flux was at least five times the
average noise level in each frame. To increase the signal-to-noise at
larger radii, we also computed color maps in which the original image
was block averaged by $2\times2$ or $4\times4$ pixels. Each of these
maps was also clipped at its 5 $\sigma$ level and sampled at the PC
pixel scale. The three maps were then combined, with each image being
weighted by its inverse blocking size. This allows one to have a
composite color map in which the bright center is shown at full
resolution and the outer, low-surface brightness regions (which were
clipped in the full resolution map) are seen at lower resolution.
This is similar to an unsharp mask technique.

\subsection{VLA Observations and Data Reduction}
We observed the galaxy with the VLA in A-configuration at 8.46 GHz and
15 GHz on 1999 August 01 in snapshot mode for 5 mins, and at 4.85 GHz
on 1998 May 21 for 10 mins. We observed a phase calibrator at the
beginning and end of the scan and 3C 48 as a flux density
calibrator. Using the AIPS software, the data were  
self-calibrated and maps were produced.

\section{Results}
\subsection{Radio Map}
A slightly super-resolved map of TXS 2226-184 at 8.46 GHz using a
circular restoring beam of 0.2\arcsec{} is shown in Figure
\ref{radiomap} (bottom) where we have subtracted the central point
source to show the extended emission more clearly. The source is
resolved with a peak flux density of 15 mJy and a total flux density
of 23 mJy. The emission is elongated in PA $-37^\circ$ towards the NW
and in PA $146^\circ$ towards the SE. No further extended emission was
detected in our maps. This is also true for lower-resolution maps (VLA
C- \& B-configuration) at 5 and 8 where the flux densities agree with
ours (Golub \& Braatz 1998). The total fluxes at 4.85 GHz and 14.94
GHz are 37 and 13 mJy respectively. At these frequencies the source is
extended in the same direction as at 8.46 GHz. If we compare our total
flux densities with the flux density the galaxy had in the Texas
survey at 365 MHz (198 mJy; Douglas et al.~1996), we find the spectrum
to be steepening from $\alpha=-0.65$ ($S_\nu\propto\nu^\alpha$)
between 365 MHz and 4.85 GHz to $\alpha=-1$ between 8.46 GHz and 14.94
GHz. Because of the compact structure this steepening is most likely
not due to resolution effects.  The position of the central radio
component is $\alpha=22^h26^m30\fs07$,
$\delta=-18\arcdeg26^\prime09\farcs6$ (B1950).

\subsection{HST Images}
Our HST images are shown in Figure~\ref{images}. The continuum map,
which is the combination of the red and green filters used also for
off-band subtraction, reveals a highly elongated galaxy along PA
$55^\circ$. The inner region (1\arcsec{} diameter) is bisected by a
dark band, presumably a nuclear dust lane. We have fitted an
elliptical Gaussian function 
to the inner region to locate the centroid of the
continuum emission. The centroid thus found is marked with a cross in
Fig.~\ref{images} and we shall refer to this position as the
``nucleus'' of the galaxy. It is in the middle of the supposed dust
lane. The presence of this dust lane is further strengthened by the  
color map, which shows a region of high reddening along PA 60$^\circ$
extending roughly 1\arcsec{} across the nucleus. We also see higher
reddening on the NW side of the galaxy than on the SE which, for a
disk galaxy, would indicate that the NW side is the nearer side of the
galaxy disk (Hubble 1943).

The H$\alpha$+[\ion{N}{2}] map shows a highly elongated structure
roughly along PA $-40^\circ\pm5^\circ$, i.e. in the same direction as
the radio emission, with a bright spot 0\farcs2 NW of the supposed
nucleus. The emission extends further towards the SE, with a broad,
``wiggly'' structure near the nucleus and a ``plume'' 1\farcs5 from
the nucleus. As in the continuum image, the adopted nucleus is not
very bright in H$\alpha$+[\ion{N}{2}], presumably because of
obscuration by the dust lane.

The adopted nucleus in the HST images is within 1\farcs5---the typical
error in absolute HST astro\-metry---of the radio nucleus. Therefore
we have assumed that the optical and radio nuclei coincide and shifted
the HST images accordingly (see Falcke et al.~1998 for a discussion of
VLA/HST registration and errors). The coordinates given in
Fig.~\ref{images} are after this shift has been performed.

In the larger field of view of all four WFPC2 chips, we find a number
of faint, extended sources around TXS2226-184 which are probably
galaxies. In particular, there is a highly elongated galaxy only
17\farcs2 SW (PA = --120$^\circ$) of the nucleus of TXS2226-184 at
$\alpha=22^h26^m29\fs0$, $\delta=-18\arcdeg26^\prime18\farcs3$
(B1950).

\begin{figure*}[htb]
\setcounter{figure}{1}
\centerline{\psfig{figure=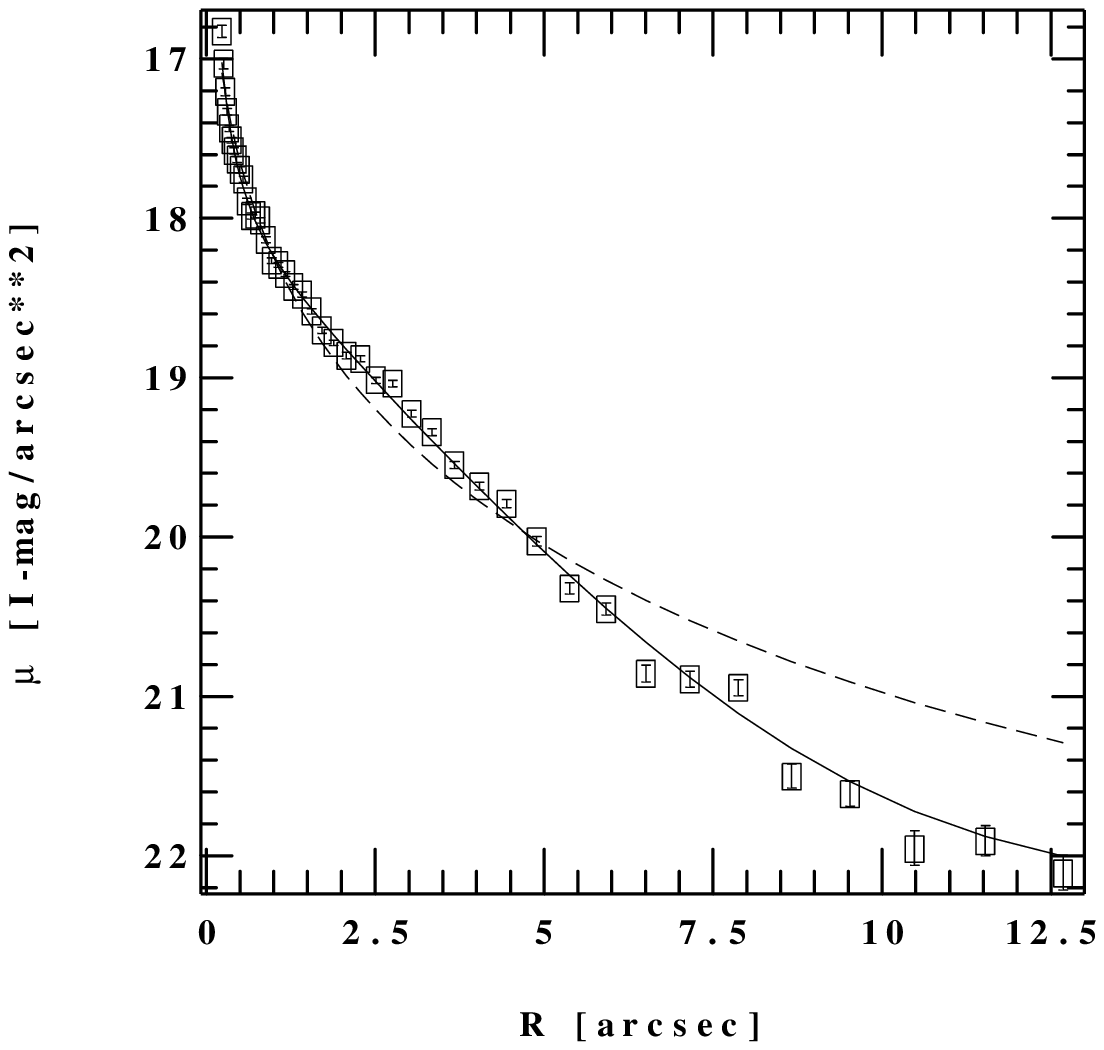,width=0.65\textwidth,bbllx=3.5cm,bburx=14.6cm,bblly=8.8cm,bbury=19.3cm}}
\caption[]{\label{fit}Surface brightness $\mu$ of TXS2226-184 in the F814W
filter (I band) as a function of the semi-major axis in
arcseconds. The solid line is a disk galaxy fit with both bulge and
disk components, as described in the text. The dashed line is a fit
with only a bulge component.}
\end{figure*}

\subsection{Isophotes and Radial Profile Fitting}
We have fitted elliptical isophotes to the red continuum image of the
galaxy ignoring the innermost few pixels which are heavily affected by
the dust lane. The center was fixed at the adopted nucleus (see
Sec. 3.2).  The ellipticity is close to zero at $R\simeq0\farcs5$,
below which it is strongly affected by the dust lane, and approaches a
constant value of around 0.6 beyond $R\ga3$\arcsec{}. Similarly, the
PA of the semi-major axis changes rapidly from $140^\circ$ to a value
of 65$^\circ$ at 0\farcs5, and stays essentially constant (at
50\arcdeg-60\arcdeg) at larger radii.  The colors are relatively red
in the inner region, dropping from V--I$\sim$1.65 to around 1.35 at
the outer isophotes.

Figure \ref{fit} shows the azimuthally averaged surface brightness of
the isophotes as a function of $R$. This profile was fitted in IRAF
with a) an exponential disk profile,

\begin{equation}
S_{\rm disk}  = S_0\cdot\exp\left(-{R\over R_0}\right),
\end{equation}
plus a bulge component (de Vaucouleurs 1948),

\begin{equation}
S_{\rm bulge} = S_{\rm e}\cdot\exp\left(-7.688\cdot\left(\left({R_{\rm
SMA}\over R_{\rm e}}\right)^{1/4} - 1\right)\right),
\end{equation}
to represent a spiral or S0 galaxy, and b) with a bulge component
(Eq.~2) only to represent an elliptical galaxy. While the fitting was
done using surface brightness, $S$, weighted by the inverse errors, we
give the results in the more conventional form of surface brightess
$\mu$ (in mag arcsec$^{-2}$).

For the disk + bulge model (a) we obtained a good fit (reduced
$\chi^2=0.86$) with the parameters $\mu_0=18.0$ mag arcsec$^{-2}$,
$R_0=2\farcs4$ (1.1 kpc), $\mu_{\rm e}=19.7$ mag arcsec$^{-2}$, and
$R_{\rm e}=0\farcs6$ (0.29 kpc). For a bulge component only (b),
i.e. an elliptical galaxy profile, the fit is much worse (reduced
$\chi^2=4.9$) and at large radii lies consistently above the data
(Fig.~\ref{fit}). The parameters we get here are $\mu_{\rm e}=22.8$
mag arcsec$^{-2}$ and $R_{\rm e}=22\farcs7$ (10.5 kpc).  The results
clearly favor a spiral over an elliptical galaxy. The ellipticity of
TXS2226-184 ($e=1-b/a=0.61$ at 2\farcs7$<R<$6\farcs0) indicates an
inclination of the galaxy to the line of sight of 70$^\circ$ (using
$i=\arcsin{\sqrt{(1-(b/a)^2)/0.96}}$, e.g.~Whittle 1992). The details
of the fitting depend somewhat on how much of the inner region is
excluded, while the preference of a disk + bulge model over a
bulge-only model does not.


The difference between the magnitudes of the integrated bulge and the
galaxy as a whole in our spiral galaxy model (see Simien \& De
Vaucouleurs 1986) is $\Delta m_{\rm t}=1.9$ if we integrate along
elliptical isophotes with $e=0.61$. To correct for the inclination
dependent absorption (e.g. Tully et al.~1998) we would have to add
$\sim$0.5 mag to obtain the face-on value of this difference. Figure 2
and Eq.~4 of Simien \& De Vaucouleurs (1986) then would formally
indicate that TXS2228-184 is probably an Sb/c (RC2 Hubble type
$T=$4-5). However, this determination of the relative bulge luminosity
and the Hubble type classification is very uncertain. Still, our data
should be good enough to indicate that TXS2228-184 is later than S0.
The fact that we are measuring at I (Simien \& De Vaucouleurs use B)
strengthens this point, since one would expect the bulge to be more
prominent relative to the disk at I than at B. If we integrate our
surface luminosity profile to infinity the total I magnitude of disk
and bulge is 15.1 mag. The uncertainty in the cut-off radius due to a
low signal-to-noise in the outer isophotes may allow an increase of
this value by up to 0.4 mag.

\section{Discussion \& Summary}
Koekemoer et al.~(1995) have classified this galaxy as an elliptical
or S0 and speculated whether the unusually broad line-width of the
megamaser emission seen in this galaxy and in NGC1052 might be typical
of elliptical galaxies. Our HST images reveal that TXS2226-184 is
almost certainly not an elliptical, so NGC1052 is the only known
megamaser in an elliptical galaxy (Braatz, Wilson \& Henkel 1994).  On
the other hand, the high inclination of TXS2226-184 strengthens the
tentative conclusion of Braatz, Wilson, \& Henkel (1997) that
megamasers are preferentially found in highly inclined galaxies. Six
out of fourteen spirals in their detected megamaser sample have now an
inclination $i>69^\circ$. This excess suggests that nuclear and large
scale dust disks in many active spiral galaxies are indeed related.

The NLR in TXS2226-184 is very elongated and reminiscent of the
jet-like NLR seen in many Seyfert galaxies, as imaged by HST
(e.g.~Capetti et al.~1996; Falcke et al.~1998).  These gaseous
structures are believed to be produced in the interaction between
outflowing radio ejecta and the ISM (e.g. Falcke et al.~1998; Ferruit
et al.~1999). The fact that our radio map is elongated along exactly
the same direction as the NLR supports this view.

In addition to the NLR and radio jet, we find a dust lane in the
nucleus which aligns with the galaxy major axis and presumably
represents its normal interstellar medium. The elongation of the NLR
and the radio source perpendicular to the NE-SW dust lane suggests
that the nuclear accretion disk and the obscuring torus are more or
less coplanar with the stellar disk in TXS2226-184. Preliminary
results of VLBA observations of the masers in this galaxy indeed seem
to roughly show a NE-SW orientation along PA 20$^\circ$ (Greenhill
1999).  How to interpret this structure and whether this indicates a
warp in the gas disk going from tens of pc to pc scales is unclear at
present. Further VLBI observations of masers and the continuum in this
and other maser sources together with HST observations of the host
galaxies could help to clarify the nature of the obscuring
torus/masing disk and its connection to the large scale molecular gas
structure of the AGN host galaxy.

\acknowledgements We thank Stacy McGaugh for helpful discussions on
galaxy classfications and Lincoln Greenhill for providing informations
on unpublished VLBA observations of TXS2226-184. This research was
supported by NASA under grant NAG8-1027 and HST GO 7278 and by NATO
grant SA.5-2-05 (GRG 960086)318/96. HF is supported by DFG grant
358/1-1\&2.


%

\clearpage
\begin{figure*}
\setcounter{figure}{0}
\centerline{\psfig{figure=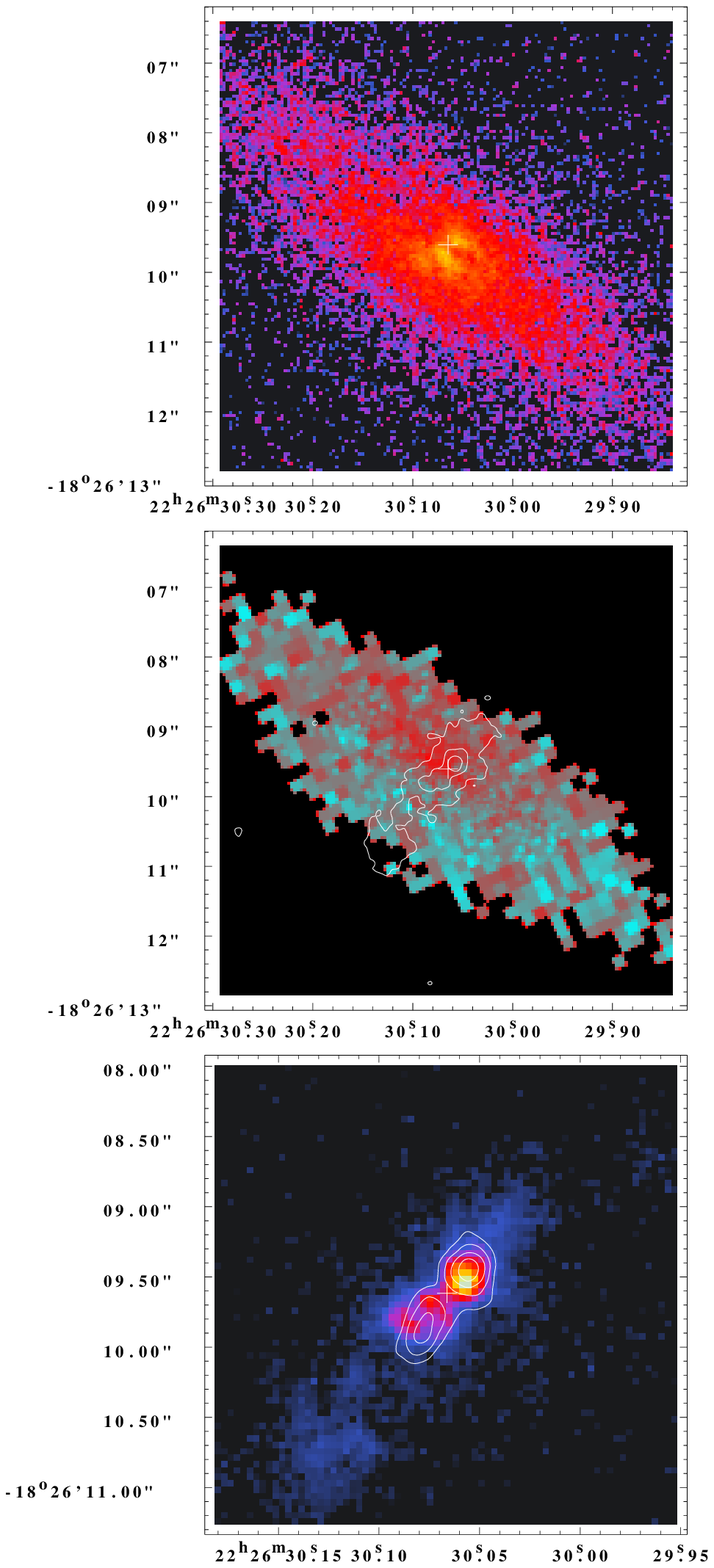,height=0.8\textheight,bbllx=5.6cm,bburx=15.3cm,bblly=3.4cm,bbury=24.9cm}}
\caption[]{\label{images}\label{radiomap} Top: continuum map
obtained by averaging the red and green images taken with the
Planetary Camera (0\farcs0455 pixel size). The centroid of the
continuum in the inner part of the galaxy (see text) is marked here
and in the following panels by a cross, and the B1950 coordinates are
from the VLA astrometry (assuming the radio nucleus and the optical
centroid are coincident). Middle: color map obtained by dividing the
green by the red continuum image (same spatial scale as top).  The
flux density ratio ranges from 0.4 (red colors) to 1.5 (blue colors)
which roughly corresponds to V--I colors ranging from 2.2 to 0.8. The
gray areas are around V--I$\sim$1.3.  Contours overlaid are of the
H$\alpha$+[\ion{N}{2}] image (bottom).  Bottom: continuum subtracted
H$\alpha$+[\ion{N}{2}] image of TXS2226-184. The
H$\alpha$+[\ion{N}{2}] flux in a rectangular
1.7\arcsec{}$\times$3.2\arcsec{} aperture is $2.5\times10^{-14}$ erg
s$^{-1}$ cm$^{-2}$ and the intensity scale is proportional to the
square root of the brightness. Contours overlaid are of the 8.46 GHz
VLA radio continuum (contours starting at 0.3 mJy and increasing by
factors of $\sqrt{2}$). We have subtracted the central point source
from the radio map to show the extended structure more clearly.}
\end{figure*}

\end{document}